\PassOptionsToPackage{table}{xcolor}
\documentclass[unnumsec,webpdf,contemporary,large]{oup-authoring-template}%

\graphicspath{{Fig/}}

\theoremstyle{thmstyleone}%
\theoremstyle{thmstyletwo}%
\theoremstyle{thmstylethree}%

\newcommand{\f}[1]{\textbf{#1}}
\newcommand{\s}[1]{\underline{#1}}
\newcommand{\std}[1]{{\scriptsize $\pm$#1}}
\newcommand{\method}[0]{INDIGENA}
\newcommand{\cc}{\cellcolor{white}}
\newcommand{\n}[1]{\ensuremath{\mathbf{#1}}}
\newcommand{\kge}{\ensuremath{f(g,r,d)}}
\newcommand{\sime}{\ensuremath{sim^e(g,d)}}
\newcommand{\simbma}{\ensuremath{sim_{BMA}^e(P_g,P_d)}}
\newcommand{\simbmm}{\ensuremath{sim_{BMM}^e(P_g,P_d)}}
\usepackage{todonotes}
\usepackage{booktabs}
\usepackage{adjustbox}
\usepackage{amsmath,amssymb}
\usepackage{xcolor}

\definecolor{myblue}{HTML}{35beff}
\definecolor{mygreen}{HTML}{7edb73}
\definecolor{myred}{HTML}{ffa2ac}
\definecolor{myorange}{HTML}{ffbc35}

\newcommand{\bluecircle}{{\color{myblue}\raisebox{-0.3ex}{\scalebox{2}{$\bullet$}}}}
\newcommand{\greencircle}{{\color{mygreen}\raisebox{-0.3ex}{\scalebox{2}{$\bullet$}}}}
\newcommand{\redcircle}{{\color{myred}\raisebox{-0.3ex}{\scalebox{2}{$\bullet$}}}}
\newcommand{\orangecircle}{{\color{myorange}\raisebox{-0.3ex}{\scalebox{2}{$\bullet$}}}}

\renewcommand{\cite}{\citep}

\begin{document}

\journaltitle{Preprint}
\copyrightyear{2026}
\appnotes{Paper}

\firstpage{1}


\title[\method{}: inductive disease--gene association prediction]{\method{}: inductive prediction of disease--gene associations using phenotype ontologies}

\author[1,2,3]{Fernando Zhapa-Camacho\ORCID{0000-0002-0710-2259}}
\author[1,2,3,$\ast$]{Robert Hoehndorf\ORCID{0000-0001-8149-5890}}

\authormark{Fernando Zhapa-Camacho et al.}

\address[1]{\orgdiv{Computer, Electrical and Mathematical Sciences \&
    Engineering Division}, \orgname{King Abdullah University of Science and
    Technology}, \orgaddress{\street{4700 KAUST}, \postcode{23955},
    \state{Thuwal}, \country{Saudi Arabia}}}


\address[2]{\orgdiv{KAUST Center of Excellence for Smart Health
    (KCSH)}, \orgname{King Abdullah University of Science and
    Technology}, \orgaddress{\street{4700 King Abdullah University of
      Science and Technology}, \state{Thuwal}, \country{Saudi
      Arabia}}}

\address[3]{\orgdiv{KAUST Center of Excellence for Generative AI},
  \orgname{King Abdullah University of Science and Technology},
  \orgaddress{\street{4700 King Abdullah University of Science and
      Technology}, \state{Thuwal}, \country{Saudi Arabia}}}

\corresp[$\ast$]{Corresponding author.
  \href{email:robert.hoehndorf@kaust.edu.sa}{robert.hoehndorf@kaust.edu.sa}}




\abstract{ \textbf{Motivation:} Predicting gene--disease associations
  (GDAs) is the problem to determine which gene is associated with a
  disease. GDA prediction can be framed as a ranking problem where
  genes are ranked for a query disease, based on features such as
  phenotypic similarity. By describing phenotypes using phenotype
  ontologies, ontology-based semantic similarity measures can be
  used. However, traditional semantic similarity measures use only the
  ontology taxonomy. Recent methods based on ontology embeddings
  compare phenotypes in latent space; these methods can use all
  ontology axioms as well as a supervised signal, but are inherently
  transductive, i.e., query entities must already be known at the time
  of learning embeddings, and therefore these methods do not
  generalize to novel diseases (sets of phenotypes) at inference time.
  \\
  \textbf{Results:} We developed \method{}, an \textbf{in}ductive
  \textbf{di}sease--\textbf{gen}e \textbf{a}ssociation method for
  ranking genes based on a set of phenotypes. Our method first uses a
  graph projection to map axioms from phenotype ontologies to a graph
  structure, and then uses graph embeddings to create latent
  representations of phenotypes. We use an explicit aggregation
  strategy to combine phenotype embeddings into representations of
  genes or diseases, allowing us to generalize to novel sets of
  phenotypes. We also develop a method to make the phenotype
  embeddings and the similarity measure task-specific by including a
  supervised signal from known gene--disease associations. We apply
  our method to mouse models of human disease and demonstrate that we
  can significantly improve over the inductive semantic similarity
  baseline measures, and reach a performance similar to transductive
  methods for predicting gene--disease associations while being more
  general.
  \\
  \textbf{Availability and Implementation:}
  \url{https://github.com/bio-ontology-research-group/indigena}
  \\
  \textbf{Contact:}
  \href{robert.hoehndorf@kaust.edu.sa}{robert.hoehndorf@kaust.edu.sa}
  \\
} 


\maketitle

\section{Introduction}


Gene--disease associations (GDAs) for Mendelian diseases can be
identified through multiple computational approaches. Mendelian
diseases, characterized by single-gene mutations following Mendelian
inheritance patterns, are predominantly rare disorders affecting fewer
than 1 in 2,000 individuals \cite{umair2023undiagnosed}. Computational
methods for identifying these associations include: (1)
guilt-by-association methods that leverage biological networks where
genes with similar network properties to known disease genes are
implicated in similar diseases \cite{kohler2008walking} (2) phenotype
similarity to databases of patients or diseases that connects patients
with similar phenotypic profiles to known genetic causes
\cite{kohler2009clinical} and (3) phenotype similarity to model
organisms, which provides substantially more data for inference
\cite{hoehndorf2011phenomenet}. This last approach is particularly
useful as wet lab validation of gene--disease associations remains
time-consuming and expensive \cite{nunes2021predicting}, while
computational methods can efficiently leverage large gene--phenotype
and disease--phenotype datasets \cite{yuan2022evaluation}. For rare
Mendelian diseases, these approaches are crucial for advancing
diagnosis and treatment options for affected patients.

Phenotypes are recorded using standardized ontologies that enable
computational analysis. Human phenotypes are described using the Human
Phenotype Ontology (HPO) \cite{gargano2024human}, while mouse
phenotypes are captured in the Mammalian Phenotype Ontology (MP)
\cite{smith2009mammalian}.  These species-specific ontologies
structure phenotypes hierarchically with formal logical
definitions. Cross-species ontologies such as UPheno
\cite{matentzoglu2024unified} and PhenomeNET
\cite{hoehndorf2011phenomenet} facilitate comparisons between human
and mouse phenotypes by relating classes of phenotypes in different
species axiomatically and thereby making them comparable.

Phenotypes are then used to predict gene--disease associations through
a ranked retrieval approach. This process involves using a phenotype
similarity measure (a semantic similarity measure) to query a database
of genotype--phenotype associations, e.g., from the Online Mendelian
Inheritance in Men (OMIM) \cite{amberger2015omim} or the Mouse Genome
Informatics (MGI) \cite{baldarelli2024mouse} databases. Genotypes,
usually representing loss of function of one or two alleles of a gene,
are then ranked based on their phenotype similarity to the query
disease. This ranking enables prioritization of candidate genes that
are most likely to be causally related to a disease
\cite{gkoutos2018anatomy}. The effectiveness of this approach relies
on the accuracy of the phenotype similarity measure and how complete
the underlying phenotype data is.

Traditional semantic similarity measures for phenotype comparison are
typically hand-crafted and can generalize to novel phenotype sets for
querying \cite{Harispe_2015}. Examples include Resnik's information
content-based measure \cite{resnik1995using} combined with the Best
Match Average (BMA) approach \cite{pesquita2009semantic} for combining
multiple comparisons.

These measures have been successfully applied to the gene--disease
association prediction tasks \cite{putman2024monarch,
  alghamdi2022contribution, hoehndorf2011phenomenet,
  smedley2013phenodigm, chen2012mousefinder}. However, semantic
similarity measures primarily rely on the phenotype ontology's
hierarchical structure and do not consider other axioms between
phenotypes. Furthermore, because semantic similarity measures are
hand-crafted, they do not adapt to the data or task of predicting
gene--disease associations.  More recent work has applied machine
learning to generate embeddings of phenotypes, genes, and
diseases. These knowledge graph or ontology embeddings
\cite{chen2025ontology} learn latent representations of single
phenotypes or sets of phenotypes, which can then be used either
through a vector similarity measure to perform ranked retrieval, or
using a supervised method like a learning-to-rank approach with a
neural network \cite{chen2021predicting}.  Supervised embedding-based
methods applied to the task of predicting gene--disease associations
based on phenotype similarity include Onto2Vec
\cite{smaili2018onto2vec} and OPA2Vec \cite{smaili2019opa2vec}, DL2Vec
\cite{owl2vecstar}, OWL2Vec* \cite{owl2vecstar}, and SmuDGE
\cite{alshahrani2018semantic}.

Embedding-based approaches are inherently transductive. Transductive
learning requires that all entities (diseases, genes) that will be
used during inference must already be available during the training
phase. This means that these models cannot generalize to previously
unseen diseases without complete retraining, thereby limiting their
applicability to patients with a previously known
disease. Furthermore, the embedding methods that were applied to the
task of predicting gene--disease associations only actually improve
predictive performance over traditional semantic similarity measures
when they incorporate a supervised signal --- known gene--disease
associations --- during training. This supervised signal may introduce
bias as the model can ``memorize'' known associations rather than
learning generalizable patterns from phenotype data alone
\cite{alghamdi2022contribution}, or predict entirely based on the
number of times a certain disease or gene was seen during training
\cite{smaili2019opa2vec}.

The limitations of transductive approaches extend beyond basic
gene--disease association prediction to variant prioritization
applications. Systems such as Exomiser \cite{robinson2014improved} or
EmbedPVP \cite{althagafi2024prioritizing} combine phenotype-based
gene--disease association prediction with variant pathogenicity
measures to prioritize potentially causal variants in clinical
settings. For such applications, the ability to make inductive
predictions --- generalizing to novel patients with previously unseen
combinations of phenotypes --- is even more critical. Patients with
rare or previously uncharacterized genetic conditions cannot benefit
from approaches that require prior knowledge of their specific disease
during model training. While systems like Exomiser use semantic
similarity, recent embedding-based methods like EmbedPVP
\cite{althagafi2024prioritizing} use a transductive method for
predicting gene--disease associations; therefore, while they show a
higher predictive performance than methods based on semantic
similarity, they are more limited in their application. On the other
hand, attempts to extend embedding-based methods to the inductive
setting \cite{althagafi2024prioritizing, safana-thesis} resulted in
predictive performance that did not reach or exceed the use of
classical semantic similarity measures.

We developed \method{}, a fully \textbf{in}ductive method for
\textbf{di}sease--\textbf{gen}e \textbf{a}ssociation prediction based
on ontology embeddings while retaining a supervised learning
component. Our approach enables ranking genes based on phenotype
similarity without requiring the test diseases (sets of phenotypes) to
be present during training. We find that \method{} outperforms
traditional semantic similarity measures while maintaining the ability
to generalize to previously unseen diseases.

Our main contributions include: (1) an inductive framework for
gene--disease association prediction that generalizes to new
combinations of phenotypes; and (2) an empirical validation of our
method's effectiveness compared to established semantic similarity
measures.  We make our code available as Free Software at
\url{https://github.com/bio-ontology-research-group/indigena}.

\section{Materials and Methods}\label{sec2}

\subsection{Datasets}\label{subsec1}
We obtained gene--phenotype associations from the MGI database
\cite{baldarelli2024mouse}. Specifically, we used the file {\tt
  MGI\_GenePheno.rpt}, downloaded from the Mouse Genome Informatics
Database on August 20th, 2025. From this file we extracted gene
identifiers and their corresponding phenotypes, encoded with the MP.

For disease--phenotype associations, we used the file {\tt
  phenotype.hpoa} from the Human Phenotype Ontology database
\cite{gargano2024human}, downloaded on August 20th, 2025.

We used the UPheno cross-species phenotype ontology
\cite{matentzoglu2024unified}, downloaded from Github
~\footnote{\url{https://github.com/obophenotype/upheno-dev/releases/tag/v2025-07-21}}
with release date July 21st, 2025. For all phenotype
associations, we ensured that the phenotypes exist in the UPheno
ontology, otherwise we omit the phenotype association.

To evaluate our ability to identify gene--disease associations, we
used the file {\tt MGI\_Geno\_DiseaseDO.rpt} from the Mouse Genome
Informatics Database \cite{baldarelli2024mouse}, downloaded on July
20th, 2025.


\subsection{Ontology preprocessing and graph projection}

We define an ontology as the tuple $\mathcal{O} = (\Sigma, Ax)$, where
$\Sigma = (\mathbf{C}, \mathbf{R}, \mathbf{I})$ provides the signature
of the ontology ($\mathbf{C}$ is a set of class names, $\mathbf{R}$ is
a set of role names, and $\mathbf{I}$ is a set of individual names)
and $Ax$ provides the set of axioms over
$\Sigma$. 
We use the axioms in the UPheno ontology and added new axioms
representing gene--phenotype, disease--phenotype and gene--disease
associations. For example, for an association between a gene $g_i$ and
phenotype $p_j$, we created the axiom
$g_i \sqsubseteq \exists \mbox{has\_phenotype}. p_j$. 
Similarly, disease--phenotypes associations were transformed to axioms
$d_i \sqsubseteq \exists \mbox{has\_symptom}. p_j$ and disease--gene
associations 
as axioms $d_i \sqsubseteq \exists \mbox{associated\_with}. g_j$. A
graph projection maps $\mathcal{O}$ into a graph $\mathcal{G}$
following a specific set of rules~\cite{graph_projections}. We
projected UPheno and its extensions into graphs following the
projection rules designed by OWL2Vec*~\cite{owl2vecstar} and shown in
Supplementary Table 1.

\begin{figure*}[htbp]
  \centering
  \includegraphics[width=0.9\textwidth]{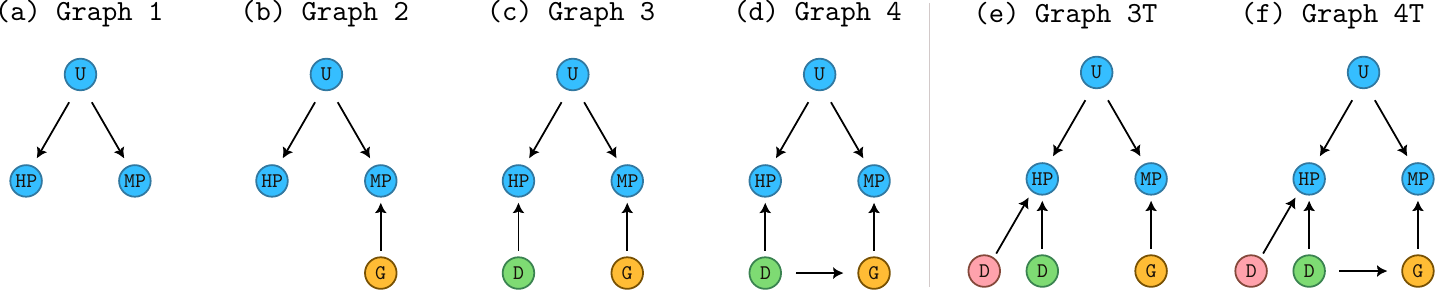}
  \caption{Different graph structures. Nodes \bluecircle{} represent
    UPheno entities, \orangecircle{} represent genes, \greencircle{}
    represent training diseases and \redcircle{} represent testing
    diseases. (a) Graph 1 is the original UPheno graph
    representation. (b) Graph 2 includes gene--phenotype
    associations. (c) Graph 3 includes disease--phenotype
    associations. (d) Graph 4 includes gene--disease associations. (e)
    and (f) are transductive variations where testing diseases have
    been added to the graph linking them to their phenotypes.}
  \label{fig:graphs}
\end{figure*}

To evaluate the impact of different information sources on prediction
performance, we constructed four distinct datasets with increasing
amounts of information. 
Graph 1 contains only the UPheno ontology, with its phenotype names
for humans and mice. This baseline graph includes the hierarchical
relationships between phenotype terms and the mappings between human
and mouse phenotypes, but no gene or disease annotations
(Figure~\ref{fig:graphs}a).
Graph 2 extends Graph 1 by adding gene--phenotype associations,
connecting MGI gene identifiers to their associated MP terms. The
phenotypes associated to a gene are those observed when the mouse
genes are mutated (Figure~\ref{fig:graphs}b).
Graph 3 further extends Graph 2 by adding disease--phenotype
associations, connecting OMIM disease identifiers to their associated
HPO phenotype terms (Figure~\ref{fig:graphs}c).
Finally, graph 4 contains a supervised signal and extends Graph 3 by
adding known gene--disease associations between MGI genes and OMIM
diseases (i.e., mouse models of human disease. This graph contains the
complete information set, including the ground truth associations that
we used for ontology embedding learning (Figure~\ref{fig:graphs}d).

Graphs 3T and 4T (Figure~\ref{fig:graphs}e,f) are transductive
variations, where test diseases are added as nodes to the training
graph with their phenotype associations only (not the gene--disease
associations). We use this graph to establish a transductive baseline
for ontology embedding methods.

These four graph structures allow us to systematically evaluate how
additional information affects the model's ability to predict
gene-disease associations. By comparing performance across these
structures, we can determine whether the inclusion of gene--phenotype,
disease--phenotype, or direct gene--disease associations impacts
prediction accuracy.

To ensure our method is truly inductive and can generalize to
previously unseen diseases, we implemented a 10-fold cross-validation
strategy based on disease splits. For Graphs 3 and 4, which contain
disease entities, the disease set was randomly partitioned into 10
equally-sized subsets. In each fold, 90\% of diseases were used for
training and 10\% were held out for testing. This procedure guarantees
that the diseases used for evaluation were never seen during the
training phase, and thereby validates the model's ability to make
predictions on novel diseases based solely on their phenotypic
profiles.

\subsection{Graph Embedding Methods}

We use several knowledge graph embedding
methods~\cite{wang2017knowledge} for our experiments, specifically
TransE \cite{bordes2013translating}, TransH \cite{Wang_2014}, TransD
\cite{ji2015knowledge} and ConvKB~\cite{nguyen-etal-2018-novel}. Each
embedding method captures different entity and relation patterns from
the knowledge graph into the embedding space depending on the scoring
function $f(h,r,t)$ (Table~\ref{tab:kge_models}).  TransE models
relationships as translations in the embedding space. TransH projects
entities into relation-specific hyperplanes, where relations are
interpreted as translations. TransD introduces entity-specific
projection vectors, that allow projecting head and tail entities
differently. While the Trans\{E,H,D\} methods provide a geometric
intepretation of embeddings, we also use ConvKB, which employs a
convolutional neural network on the concatenated embeddings of
entities and relations. The original ConvKB model is initialized with
TransE embeddings. Furthermore, we explore the impact of initializing
ConvKB with TransD embeddings, which we name ConvKB-D.

\begin{table*}[htbp]
  \centering
  \caption{Summary of Knowledge Graph Embedding Models}
  \label{tab:kge_models}
  \begin{tabular}{llll}
    \toprule
    Method & Ent. Embedding & Rel. Embedding & Scoring function \\
    \midrule
TransE & $\n{h}, \n{t} \in \mathbb{R}^d$ & $\n{r} \in \mathbb{R}^d$ & $f(h,r,t)= -\|\n{h} + \n{r} - \n{t}\|_{1/2}$ \\
    TransH & $\n{h}, \n{t} \in \mathbb{R}^d$ & $\n{r}, \n{w}_r \in \mathbb{R}^d$ & $f(h,r,t)=-\|(\n{h} - \n{w}_r^\top\n{h}\n{w}_t) + \n{r} - (\n{t} - \n{w}_r^\top\n{t}\n{w}_r)\|_2^2$\\
TransD & $\n{h}, \n{t}, \n{w}_h, \n{w}_t \in \mathbb{R}^d$ & $\n{r}, \n{w}_r \in \mathbb{R}^k$ & $f(h,r,t)=-\|(\n{w}_r\n{w}_h^\top + \n{I})\n{h} + \n{r} - (\n{w}_r\n{w}_t^\top + \n{I})\n{t}\|_2^2$ \\
    
    ConvKB & $\n{h}, \n{t} \in \mathbb{R}^d$ & $\n{r} \in \mathbb{R}^d$ & $f(h,r,t)=[\n{v}_1;\ldots;\n{v}_\tau]\cdot \n{w},\quad \n{v}_i=g(\omega_j\n{[\n{h},\n{r},\n{t}] +\n{b}} ),\quad \n{w}\in \mathbb{R}^{\tau d}$ \\
    \bottomrule
  \end{tabular}
\end{table*}

\subsection{Semantic Similarity}

Semantic similarity measures quantify the likeness of concepts based
on their meaning and relationships within an ontology
\cite{pesquita2009semantic}. For phenotype-based gene--disease
association prediction, we use two approaches: semantic similarity
measures and embedding-based similarity.

Regarding semantic similarity measures we used Resnik
\cite{resnik1995using} and Lin~\cite{linsemsim}. These measures
quantifies the similarity between two ontology pairs of terms based on
their shared information content (IC) with Lin's measure being an
variation of Resnik's. For two phenotype terms $p_1$ and $p_2$, the
similarity is defined as:
\begin{equation}
sim_{Resnik}(p_1, p_2) = IC(MICA(p_1, p_2))
\end{equation}
\begin{equation}
  sim_{Lin}(p_1, p_2) = \frac{2\cdot IC(MICA(p_1, p_2))}{IC(p_1)+IC(p_2)}
\end{equation}
where $MICA(p_1, p_2)$ is the most informative common ancestor of
$p_1$ and $p_2$ in the ontology hierarchy, and $IC(p)$ is the
information content of term $p$, calculated as:
\begin{equation}
IC(p) = -\log(P(p))
\end{equation}
with $P(p)$ representing the frequency of phenotype term $p$ in the
corpus of annotations. We used the Semantic Measures Library
\cite{Harispe_2013, Harispe_2015} to compute semantic 
similarity measures.

To compute similarity between sets of terms, we rely of two indirect
approaches (i.e., measures that first compute pairwise measures and
then combine), the Best-Match-Average (BMA) and Best-Match-Maximum
(BMM), and one direct approach (i.e., a measure that does not rely on
pairwise comparisons), SimGIC~\cite{pesquita2007evaluating}:

\begin{align}
  \label{eq:bma}
  sim_{BMA}(P_g, P_d) & = \frac{1}{2} \left(\frac{\sum_{i=1}^{n} \max_{j} sim(p_{gi}, p_{dj})}{n} \right. \nonumber \\
                      & \quad \quad \left. + \frac{\sum_{j=1}^{m} \max_{i} sim(p_{gi}, p_{dj})}{m}\right)
\end{align}
\begin{align}
  \label{eq:bmm}
  sim_{BMM}(P_g, P_d) & = \max \left(\frac{\sum_{i=1}^{n} \max_{j} sim(p_{gi}, p_{dj})}{n} \right. \nonumber \\
                      & \quad \quad \quad \left. , \frac{\sum_{j=1}^{m} \max_{i} sim(p_{gi}, p_{dj})}{m}\right)
\end{align}
\begin{equation}
  \label{eq:simgic}
  sim_{GIC}(P_g, P_d) = \frac{\sum_{p \in C^+_T(P_g)\cap C^+_T(P_d)}^{n} IC(c)}{\sum_{p \in C^+_T(P_g)\cup C^+_T(P_d)}^{n} IC(c)}
\end{equation}
where $C^+_T(X)$ is the union of ancestors of the concepts in $X$.

\subsection{Embedding-based similarity}
In order to provide an inductive prediction framework, we use the
vector representations generated by knowledge graph embedding
models. Notice that graph embedding methods are optimized with the
scoring function $f(h,r,t)$. However, we work under the assumption
that the embeddings capture similarity features during the training
process. Therefore, given two entity terms with embeddings $\vec{p_1}$
and $\vec{p_2}$, we calculate their similarity as follows:
\begin{equation}
  sim^{e}(p_1, p_2) = \sigma(\langle \vec{p_1}, \vec{p_2} \rangle)
\end{equation}
where $\langle \cdot, \cdot \rangle$ represents the dot product
between vectors and $\sigma(\cdot)$ is the sigmoid function.
For a gene $g$ and disease $d$, we can compute their similarity in two
ways: (a) using the entities directly in the expression $sim^{e}(g,d)$
or (b) using their phenotype representations with different aggregation methods such as BMA (Equation
\ref{eq:sim_emb_bma}) and BMM (Equation~\ref{eq:sim_emb_bmm}).

\begin{align}
  \label{eq:sim_emb_bma}
  sim_{BMA}^{e}(P_g, P_d) & = \frac{1}{2} \left(\frac{\sum_{i=1}^{n} \max_{j} sim^{e}(p_{gi}, p_{dj})}{n} \right. \nonumber \\
                          & \quad \quad \left. + \frac{\sum_{j=1}^{m} \max_{i} sim^{e}(p_{gi}, p_{dj})}{m}\right)
\end{align}

\begin{align}
  \label{eq:sim_emb_bmm}
  sim_{BMM}^{e}(P_g, P_d) & = \max \left(\frac{\sum_{i=1}^{n} \max_{j} sim^{e}(p_{gi}, p_{dj})}{n} \right. \nonumber \\
                          & \quad \quad \left. + \frac{\sum_{j=1}^{m} \max_{i} sim^{e}(p_{gi}, p_{dj})}{m}\right)
\end{align}

\section{Results}

\subsection{Transductive prediction of gene--disease associations}
\label{subsec:transductive}

We start analyzing the prediction capabilities of embedding methods in
a transductive setting. For this task, we use Graphs 3T and 4T, which
contain the testing diseases in the training set. Graph 4T also
includes gene--disease associations for training diseases; therefore
we say Graph 4T includes includes a \emph{supervised signal}.


We evaluated five different knowledge graph embedding methods: TransE,
TransH, TransD, ConvKB and ConvKB-D and show the results in
Table~\ref{tab:transductive}.  First, we find that introducing a
supervised signal (Graph 4T) generally performs better than using no
supervised information (Graph 3T). While there are some exceptions
(TransE), most embedding methods can take advantage of the supervised
signal.  Second, embeddings can capture similarity features during
training. 

\begin{figure*}[htbp]
  \centering
  \includegraphics[width=0.95\textwidth]{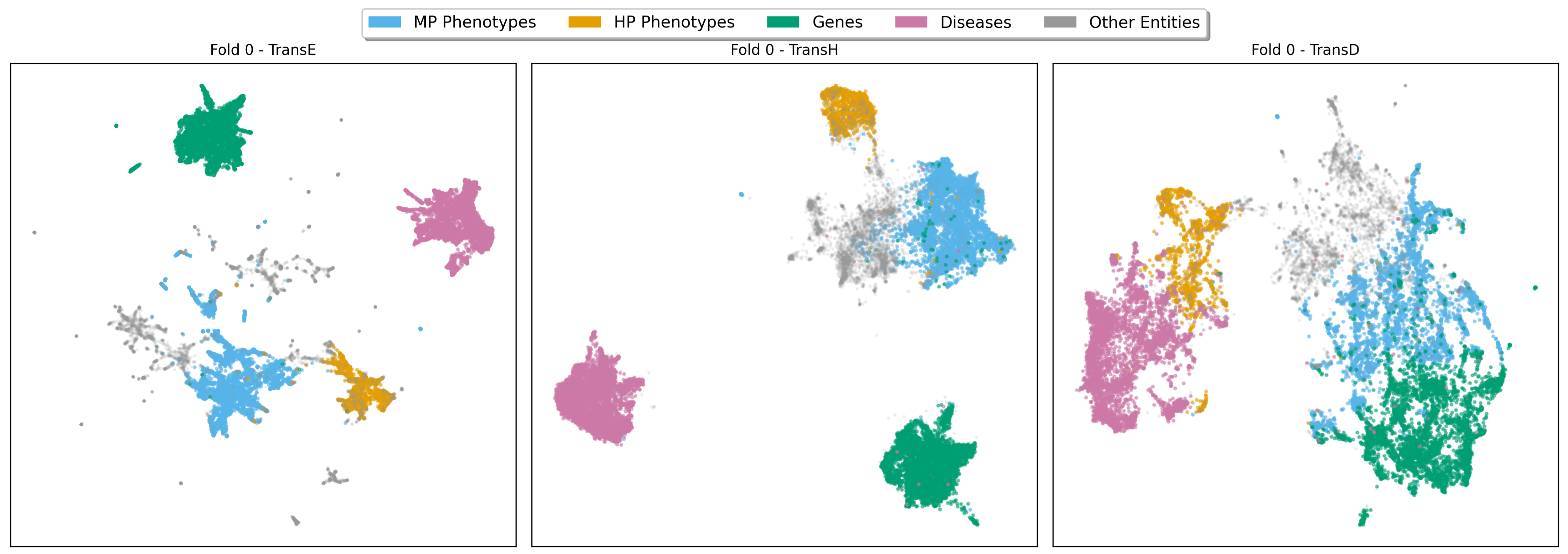}
  \caption{UMAP representation of learned embeddings for methods
    TransE, TransH and TransD on the first fold. TransE learns a
    distance function direcly, therefore, their embeddings are
    scattered across the latent space. TransH and TransD learn a
    projection function; therefore the initial latent space capture
    similarity features of embeddings.}
  \label{fig:umap}
\end{figure*}

In TransE, embeddings are optimized to predict relations using the
distance-based scoring function $\kge{}$ directly. Consequently,
evaluation metrics based on $\kge{}$ yield the best results for this
model. In contrast, TransH and TransD project entity embeddings into a
relation-specific subspace before computing the score. Because the
model relies on these projections to handle relational properties, the
base embeddings are free to capture intrinsic semantic
similarities. This explains why similarity-based metrics ($\sime{}$,
$\simbmm{}$, $\simbma{}$) outperform the standard scoring function
$\kge{}$ for TransH and TransD. Furthermore, TransD outperforms TransH
due to its more expressive projection mechanism. Figure \ref{fig:umap}
illustrates the UMAP~\cite{mcinnes2018umap-software} projections of
the learned embeddings. TransE embeddings are widely dispersed across
the latent space to satisfy various translational constraints
directly. Conversely, TransH and TransD embeddings exhibit distinct
clustering; by offloading relational complexity to the projection
step, the base embeddings effectively encode semantic similarity.

We also analyze ConvKB, which uses a convolutional neural network on
the concatenated embeddings of entities and relations. Originally,
ConvKB is initialized with pretrained TransE embeddings, and given
that TransD outperforms TransE in our analysis, we initialized ConvKB
with TransD embeddings as well, and named it ConvKB-D. ConvKB improves
TransE, which shows that the convolutional neural network enhance the
similarity features of original TransE embeddings. On the other hand,
while ConvKB-D improves TransD under $\sime{}$, its performance under
$\simbma{}$ does not change, suggesting that the similarity features
learned by TransD cannot be further improved by the convolutional
neural network of ConvKB-D.

In both types of measures, semantic similarity measures and embedding
similarity, we notice that the aggregation function BMA performs,
better than BMM; therefore we rely solely on BMA for the rest of the
experiments.

\begin{table*}[t]
\centering
\caption{Performance comparison of different gene--disease association
  prediction methods. Results are reported based on 10 folds. Graph
  embedding methods are trained using the scoring funcion \kge{} and
  evaluated under the scoring functions \kge{}, \sime{},
  \simbma{}. \textbf{Bolded} values show the best values in each
  column.  Shaded rows highlight the evaluation under the scoring
  function \simbma{}. $\downarrow$ means that lower values are better
  and $\uparrow$ means that higher values are
  better. \label{tab:transductive}}
\begin{adjustbox}{width=0.99\textwidth}
\begin{tabular}{l|c|cccc|cccc}
\toprule
  \textbf{Method} & \textbf{Scoring} & \textbf{MR} ($\downarrow$) & \textbf{MRR} ($\uparrow$) & \textbf{H@100} ($\uparrow$) & \textbf{AUC} ($\uparrow$) & \textbf{MR} ($\uparrow$) & \textbf{MRR} ($\uparrow$) & \textbf{H@100} ($\uparrow$) & \textbf{AUC} ($\uparrow$) \\
\midrule
& & \multicolumn{4}{c|}{Non-supervised signal (Graph 3T)} & \multicolumn{4}{c}{Supervised signal (Graph 4T)} \\
  \midrule
\multicolumn{9}{l}{\textit{Semantic similarity baselines}} \\
  SimGIC    & $sim_{GIC}$ & 275.40\std{30.14} & 0.09\std{0.01} & 0.50\std{0.05} & 0.82\std{0.02} & 275.40\std{30.14}  & 0.09\std{0.01} & 0.50\std{0.05} & 0.82\std{0.02}  \\
  Resnik &$sim_{BMM}$& 217.05\std{24.23} & 0.09\std{0.02} & 0.53\std{0.06} & 0.86\std{0.02} & 217.05\std{24.23}  & 0.09\std{0.02} & 0.53\std{0.06} & 0.86\std{0.02} \\
  Resnik & $sim_{BMA}$& 172.63\std{17.74} & \f{0.12}\std{0.02} & 0.63\std{0.04} & 0.89\std{0.01} & 172.63\std{17.74}  & {\f{0.12}}\std{0.02} & 0.63\std{0.04} & 0.89\std{0.01} \\
  Lin &$sim_{BMM}$& 204.65\std{15.65} & 0.08\std{0.01} & 0.53\std{0.03} & 0.87\std{0.01} & 204.65\std{15.65}  & 0.08\std{0.01} & 0.53\std{0.03} & 0.87\std{0.01} \\
  Lin & $sim_{BMA}$ & {161.62}\std{12.92} & \f{0.12}\std{0.02} & {0.65}\std{0.04} & {0.90}\std{0.01} & 161.62\std{12.92}  & {\f{0.12}}\std{0.02} & 0.65\std{0.04} & 0.90\std{0.01} \\
  \midrule
  \multicolumn{9}{l}{\textit{Knowledge graph embedding methods}} \\
  \multirow{3}{*}{TransE}   & $f(h,r,t)$          & --                    & --                     & --                 & --                 & 264.49\std{28.15}     & 0.06\std{0.01}     & 0.46\std{0.03}     & 0.83\std{0.02}\\
                            & $sim^{e}(h,t)$      & 269.50\std{15.71}     & 0.04\std{0.01}         & 0.42\std{0.04}     & 0.83\std{0.01}     & 316.39\std{29.01}     & 0.04\std{0.01}     & 0.36\std{0.05}     & 0.80\std{0.02}  \\
                            & $sim_{BMM}^{e}$     & 410.77\std{32.42}     & 0.02\std{0.01}         & 0.25\std{0.03}     & 0.73\std{0.02} & 425.02\std{33.42} & 0.02\std{0.01} & 0.22\std{0.05} & 0.72\std{0.02} \\
  \rowcolor{gray!50}
\cc                         & \cc $sim_{BMA}^{e}$ & 352.24\std{26.23}     & 0.04\std{0.01}         & 0.34\std{0.04}     & 0.77\std{0.02}     & 364.54\std{30.27}     & 0.03\std{0.01}     & 0.31\std{0.06}     & 0.76\std{0.02}  \\
  \midrule
  \multirow{3}{*}{TransH}   & $f(h,r,t)$          & --                    & --                     & --                 & --                 &   571.86\std{35.08}   & 0.01\std{0.00}     & 0.13\std{0.02}     & 0.63\std{0.02} \\
                            & $sim^{e}(h,t)$      &546.12\std{32.31}      & 0.01\std{0.00}         & 0.16\std{0.03}     & 0.65\std{0.02}     & 453.96\std{41.57}     & 0.02\std{0.01}     & 0.23\std{0.04}     & 0.71\std{0.03}  \\
                            & $sim_{BMM}^{e}$     & 348.34\std{41.66} & 0.04\std{0.01} & 0.34\std{0.04} & 0.77\std{0.03} & 301.47\std{35.63} & 0.04\std{0.02} & 0.39\std{0.06} & 0.81\std{0.02}  \\
  \rowcolor{gray!50}
  \cc                       & \cc $sim_{BMA}^{e}$ &297.57\std{50.22}      & 0.05\std{0.02}         & 0.43\std{0.06}     & 0.81\std{0.03}     & 246.58\std{29.89}     & 0.06\std{0.02}     & 0.47\std{0.06}     & 0.84\std{0.02}  \\
  \midrule
  \multirow{3}{*}{TransD}   & $f(h,r,t)$          & --                    & --                     & --                 & --                 & 308.33\std{55.85}     & 0.06\std{0.02}     & 0.42\std{0.07}     & 0.80\std{0.04} \\
                            & $sim^{e}j(h,t)$     & 271.77\std{50.32}     & 0.07\std{0.01}         & 0.48\std{0.05}     & 0.82\std{0.03}     & 234.37\std{53.68}     & 0.08\std{0.02}     & 0.51\std{0.08}     & 0.85\std{0.03} \\
                            & $sim_{BMM}^{e}$     & 176.41\std{36.73} & 0.09\std{0.01} & 0.59\std{0.06} & 0.89\std{0.02}  & 160.86\std{29.48} & 0.09\std{0.02} & 0.64\std{0.05} & 0.90\std{0.02}  \\
  \rowcolor{gray!50}
\cc                         & \cc $sim_{BMA}^{e}$ & 144.71\std{33.59}     & \f{\f{0.12}}\std{0.02} & \f{0.67}\std{0.06} & \f{0.91}\std{0.02} & 131.40\std{25.74}     & \f{0.12}\std{0.02} & \f{0.71}\std{0.05} & \f{0.92}\std{0.02}  \\
  \midrule
  \multirow{3}{*}{ConvKB}   & $f(h,r,t)$          & --                    & --                     & --                 & --                 & 188.31\std{23.48}     & 0.07\std{0.01}     & 0.56\std{0.04}     & 0.88\std{0.02}  \\
                            & $sim^{e}(h,t)$      & 248.62\std{17.19}     & 0.05\std{0.01}         & 0.43\std{0.04}     & 0.84\std{0.01}     & 270.47\std{20.70}     & 0.04\std{0.01}     & 0.41\std{0.05}     & 0.83\std{0.01} \\
                            & $sim_{BMM}^{e}$     & 373.77\std{31.75} & 0.03\std{0.01} & 0.29\std{0.03} & 0.76\std{0.02}  & 355.47\std{39.16} & 0.03\std{0.01} & 0.30\std{0.04} & 0.77\std{0.03}  \\

  \rowcolor{gray!50}
  \cc                       & \cc $sim_{BMA}^{e}$ & 308.80\std{29.73}     & 0.05\std{0.01}         & 0.38\std{0.05}     & 0.80\std{0.02}     & 286.00\std{33.65}     & 0.04\std{0.01}     & 0.41\std{0.05}     & 0.82\std{0.02}  \\
  \midrule
  \multirow{3}{*}{ConvKB-D} & $f(h,r,t)$          & --                    & --                     & --                 & --                 & 318.39\std{60.51}     & 0.07\std{0.02}     & 0.40\std{0.07}     & 0.79\std{0.04}  \\
                            & $sim^{e}(h,t)$      & 237.70\std{48.54}     & 0.08\std{0.01}         & 0.52\std{0.06}     & 0.85\std{0.03}     & 226.72\std{51.04}     & 0.08\std{0.02}     & 0.52\std{0.07}     & 0.85\std{0.03}  \\
                            & $sim_{BMM}^{e}$     & 173.06\std{36.36} & 0.09\std{0.02} & 0.59\std{0.06} & 0.89\std{0.02}  & 157.13\std{28.19} & 0.10\std{0.02} & 0.64\std{0.06} & 0.90\std{0.02}  \\
  \rowcolor{gray!50}
  \cc                       & \cc $sim_{BMA}^{e}$ & \f{142.16}\std{32.26} & 0.11\std{0.02}         & \f{0.67}\std{0.06} & \f{0.91}\std{0.02} & \f{130.63}\std{25.55} & \f{0.12}\std{0.03} & \f{0.71}\std{0.05} & \f{0.92}\std{0.02}  \\
  \bottomrule
\end{tabular}
\end{adjustbox}
\end{table*}

\subsection{Inductive Approach for Gene--Disease Association Prediction}

In clinical settings, the gene--disease association prediction task
frequently involves novel diseases or patients with previously
uncharacterized conditions. While the set of genes remains stable, new
diseases and phenotypic manifestations continue to be discovered,
particularly for rare Mendelian disorders. This presents a fundamental
limitation for transductive embedding approaches, which require
diseases to be present during the training phase to generate their
embeddings.

The key insight of our inductive approach is that, while diseases may
be novel, the phenotypes used to describe them are drawn from a
stable, predefined ontology. Semantic similarity measures are
inherently inductive because they operate on phenotypes rather than
directly on diseases or genes. Pairwise similarity between phenotypes
is aggregated using an aggregation function such as the
best-match average (BMA). 
This allows semantic similarity to be applied to any disease
characterized by known phenotypes, even if the disease itself was not
seen during training.

We extend this intuition to embedding-based approaches by computing
BMA scores  between phenotype
embeddings rather than directly comparing gene and disease
embeddings. Given a gene $g$ with phenotypes $P_g$ and a disease $d$
with phenotypes $P_d$, we calculate the BMA score based on pairwise
similarities between phenotype embeddings using \simbma{}. This
approach allows us to predict associations for any disease described
by a set of phenotypes, even if the disease itself was not included
during training.

To evaluate our inductive approach, we conducted experiments using the
TransD and ConvKB-D embedding models with four different graph
structures of increasing complexity:
\begin{itemize}
\item Graph 1 includes only the UPheno ontology with cross-species
  phenotype mappings, providing a baseline that captures hierarchical
  and cross-species phenotype relationships but contains no gene or
  disease information.

\item Graph 2 extends Graph 1 by adding gene--phenotype associations,
  enabling the model to learn relationships between genes and
  phenotypes while still lacking disease information. Note that the
  set of genes is always known and does not change, making this
  approach also inductive.

\item Graph 3 further extends Graph 2 by incorporating
  disease--phenotype associations from OMIM. This graph contains the
  complete phenotypic profiles of both genes and diseases but does not
  include known gene--disease associations. As long as the disease (or
  set of phenotypes) which is tested is not included in this graph,
  predicting gene--disease associations with Graph 3 can also be
  inductive. We generate 10 versions of this graph, in each of which
  we randomly remove 10\% of diseases; for inductive inference, we use
  the graph in which the test disease is not present.

\item Graph 4, the supervised graph, extends Graph 3 by adding known
  gene--disease associations. Similarly to Graph 3, we implemented
  10-fold cross-validation on diseases to ensure that test diseases
  are not seen during training.
\end{itemize}

Table~\ref{tab:inductive} presents the performance of our inductive
BMA-based approach across these graph structures.  In the case of
TransD, while Graph 1 and Graph 2 can accumulate more predictions in
the top 3, Graph 4 becomes better from top 10 onwards and also in
averaged metrics such as Mean Rank and AUC.
When training ConvKB-D, Graph 4 performance improves in the top 3, and
the performance in AUC is maintained.  In both methods (TransD and
ConvKB-D), for Graph 4, the inductive approach obtains AUC of $0.93$,
which is comparable to the transductive approach despite the more
challenging task of generalizing to unseen diseases, i.e., inductive
inference. This demonstrates that learning from phenotype patterns is
nearly as effective as directly learning from gene--disease
associations. Additionally, the metrics achieved for our methods
significantly improves over several standard semantic similarity
measures such as SimGIC, Resnik and Lin with different aggregation
methods such as BMA and BMM, confirming that learned phenotype
embeddings can capture more complex relationships than handcrafted
similarity measures based solely on information content, while still
retaining the ability for inductive inference.  To demonstrate
statistical significance, we conducted a Wilcoxon Signed-Rank
test comparing Lin-BMA (the best baseline) and ConvKB-D G4 across the
10 folds, obtaining a p-value of $1.178 \cdot 10^{-10}$
on the ranks produced by both
methods, indicating that the improvement is statistically significant
%
These results validate that our approach remains effective when
applied to diseases not seen during training, making it suitable for
real-world clinical applications where newly characterized diseases
(characterized as sets of phenotypes) are considered.

\begin{table*}[htbp]
  \centering
  \caption{Performance of different graph structures using TransD and
    ConvKB-D with the BMA-based approach. Results are based on 10-fold
    cross-validation for Graph 3 and Graph 4, ensuring test diseases
    were not seen during training. Bolded values indicate best
    performance for each metric and \underline{underlined} values
    indicate second-best performance. $\downarrow$ means that lower
    values are better and $\uparrow$ means that higher values are
    better.\label{tab:inductive}}
\begin{tabular}{lrrrrrrr}
    \toprule
    UPheno version & MR ($\downarrow$) & MRR ($\uparrow$) & H@1 ($\uparrow$) & H@3 ($\uparrow$) & H@10 ($\uparrow$) & H@100 ($\uparrow$) & AUC ($\uparrow$) \\
    \midrule
      SimGIC & 275.40\std{30.14} & 0.09\std{0.01} & 0.04\std{0.01} & 0.09\std{0.02} & 0.17\std{0.03} & 0.50\std{0.05} & 0.82\std{0.02}  \\
    Resnik-BMA & 172.63\std{17.74} & 0.12\std{0.02} & \s{0.05}\std{0.02} & 0.13\std{0.02} & 0.26\std{0.02} & 0.63\std{0.04} & 0.89\std{0.01}  \\
    Resnik-BMM & 217.05\std{24.23} & 0.09\std{0.02} & 0.04\std{0.01} & 0.08\std{0.02} & 0.19\std{0.02} & 0.53\std{0.06} & 0.86\std{0.02}  \\
    Lin-BMA & 161.62\std{12.92} & 0.12\std{0.02} & \s{0.05}\std{0.02} & 0.13\std{0.02} & 0.26\std{0.03} & 0.65\std{0.04} & 0.90\std{0.01}  \\
    Lin-BMM & 204.65\std{15.65} & 0.08\std{0.01} & 0.03\std{0.01} & 0.08\std{0.02} & 0.17\std{0.02} & 0.53\std{0.03} & 0.87\std{0.01}  \\

    \midrule
TransD G1 &  141.58\std{26.40} & \f{0.14}\std{0.02} & \f{0.06}\std{0.01} & \f{0.15}\std{0.02} & \s{0.29}\std{0.03} & 0.69\std{0.05} & 0.91\std{0.02}  \\
  TransD  G2 & 130.19\std{18.64} & \s{0.13}\std{0.02} & \s{0.05}\std{0.02} & \f{0.15}\std{0.02} & \s{0.29}\std{0.04} & 0.70\std{0.04} & \s{0.92}\std{0.01} \\
  TransD G3 & 150.69\std{29.52} & 0.12\std{0.02} & \s{0.05}\std{0.01} & 0.12\std{0.02} & 0.27\std{0.04} & 0.67\std{0.05} & 0.90\std{0.02} \\
  TransD G4 & \f{115.70}\std{35.17} & \s{0.13}\std{0.03} & \s{0.05}\std{0.03} & \s{0.14}\std{0.03} & \f{0.30}\std{0.04} & \f{0.74}\std{0.06} & \f{0.93}\std{0.02}  \\
  \midrule
ConvKB-D G1 & 138.60\std{25.54} & \f{0.14}\std{0.02} & \f{0.06}\std{0.01} & \f{0.15}\std{0.02} & \f{0.30}\std{0.03} & 0.69\std{0.05} & 0.91\std{0.02}  \\
ConvKB-D G2 & 129.38\std{19.08} & \s{0.13}\std{0.02} & \f{0.06}\std{0.02} & \s{0.14}\std{0.03} & 0.28\std{0.03} & \s{0.71}\std{0.04} & \s{0.92}\std{0.01}  \\
ConvKB-D G3 & 151.23\std{34.41} & 0.12\std{0.02} & 0.04\std{0.01} & 0.12\std{0.03} & 0.27\std{0.04} & 0.67\std{0.06} & 0.90\std{0.02}  \\
ConvKB-D G4 & \s{116.99}\std{42.41} & \f{0.14}\std{0.03} & \f{0.06}\std{0.02} & \s{0.14}\std{0.03} & \f{0.30}\std{0.05} & \f{0.74}\std{0.07} & \f{0.93}\std{0.03}  \\

        \bottomrule
  \end{tabular}
  \label{tab:graph_ablation}
\end{table*}

\section{Discussion}

\subsection{Inductive Gene--Disease Association  Prediction}

Inductive approaches for gene--disease association prediction are
critical for addressing the challenges of rare genetic disease
diagnosis. The majority of Mendelian diseases are rare, with new
conditions continually being characterized. Traditional transductive
embedding approaches cannot handle previously unseen diseases without
complete retraining, severely limiting their clinical utility. Our
inductive method addresses this limitation by enabling predictions for
novel diseases based solely on their phenotypic profiles.

This capability is particularly important when integrating
gene--disease association prediction into variant prioritization
\cite{althagafi2024prioritizing}. These systems combine
phenotype-based gene prioritization with variant pathogenicity metrics
to identify causative variants in patients with genetic disorders. By
adopting our inductive approach, these systems can handle patients
with previously uncharacterized diseases or novel combinations of
phenotypes without requiring prior knowledge of specific disease
entities during model training.

Our approach extends the state-of-the-art in multiple ways. First,
unlike traditional semantic similarity measures like Resnik or Lin,
which rely on hand-crafted metrics, our method leverages knowledge
graph embeddings to learn relationships between phenotypes across
species. Second, unlike previous embedding approaches that require
diseases to be present during training, our method operates at the
phenotype level, making it inherently inductive. Third, we retain the
benefits of supervised learning by incorporating gene--disease
associations during training while maintaining the ability to
generalize to unseen diseases. This balance between supervision and
induction is likely the most significant advancement over existing
approaches, and may open up possibilities for future improvements in
predicting gene--disease associations.

\subsection{Embedding model performance}

Our experimental results demonstrate that TransD consistently
outperforms other embedding models such as TransE and TransH for
gene--disease association prediction. TransD's scoring function
mechanism maps entities into entity-relation-specific spaces and
compute scores $f(h,r,t)$ in the target space. Thus, in the origin
space, base embeddings obtain similarity features that are used to be
projected to similar regions in the target space.  This property of
TransD makes it well-suited for capturing the relationships in our
graph. Additionally, training a ConvKB model with TransD features can
slightly improve the performance.

While the UPheno ontology alone (Graph 1) provides a foundation for
phenotype comparison, the addition of gene--phenotype associations
(Graph 2), disease--phenotype associations (Graph 3), and known
gene--disease associations (Graph 4) each contribute to improved
predictive performance.

We observed that performance decreases when moving from Graph 2 to
Graph 3 (adding disease phenotypes without gene--disease links) for
projection-based methods like TransD. We attribute this to the
structural differences between gene and disease annotations. Genes
are annotated with the Mammalian Phenotype Ontology (MP) while
diseases use the Human Phenotype Ontology (HPO); although linked via
UPheno, the phenotype annotations have different annotation
frequencies, granularities, and biases.

In the absence of direct gene--disease links, embedding methods like
TransD --- which project entities into relation-specific spaces ---
will segregate genes and diseases into distinct regions of the latent
space based on these annotation differences. The gene--disease
associations in Graph 4 then act as an alignment signal, forcing the
model to bridge this domain gap and map genes and diseases into a
shared manifold.

Conversely, we observe that TransE performance degrades with this
supervised signal (Table \ref{tab:transductive}).  TransE models the {\tt
  associated\_with} relation as a translation vector $r$, enforcing
the constraint $h+r \approx t$. This translation explicitly separates
the gene and disease embeddings by the magnitude of $r$ in the vector
space. Because our inference approach relies on direct embedding
similarity (where higher similarity corresponds to closer proximity),
this forced separation makes the phenotypes associated with genes and
diseases more dissimilar in the latent space, thereby degrading the
performance of similarity-based ranking.

Overall, our findings demonstrate that each layer of biological
knowledge can add a signal that the embedding models may leverage. We
also note that Graph 2 contains the information that can be used by
the semantic similarity measures, as information content is computed
over genes; graphs 3 and 4 are able to utilize more information than
available to semantic similarity measures.

Notably, the supervised model (Graph 4) achieves substantially higher
performance than unsupervised alternatives, highlighting the
importance of known gene--disease associations as a supervision
signal. The need for a supervised signal has also been shown in all
prior studies that rely on embeddings for computing gene--disease
associations. However, unlike previous supervised approaches that
sacrifice inductive capabilities, our method maintains the ability to
generalize to unseen diseases through the BMA-based phenotype
comparison.

\subsection{Comparison with Semantic Similarity Measures}

Our comparison with semantic similarity measures reveals important
insights about the strengths of embedding-based approaches. While
semantic similarity measures has been widely used due to its
simplicity and interpretability, it primarily relies on the
information content of the most informative common ancestor in the
ontology hierarchy. This approach neglects more complex relationships
and cannot adapt to the specific patterns in gene--disease association
data.

In contrast, our embedding-based approach learns latent
representations that capture both hierarchical and non-hierarchical
relationships in the ontology. The TransD model can identify axiom
patterns in how phenotypes relate to each other across species, which
may explain its superior performance. Additionally, by incorporating a
supervised signal during training, our model can learn which phenotype
patterns are most relevant for predicting gene--disease associations,
rather than relying solely on general semantic similarity.

\section{Conclusions}

Our study demonstrates that inductive, supervised gene--disease
association prediction can successfully address the limitations of
traditional approaches. By operating at the phenotype level rather
than directly at the disease level, our method can generalize to novel
diseases based solely on their phenotypic manifestations. This
capability is particularly valuable for rare disease diagnosis where
previously uncharacterized conditions regularly emerge.

The framework we have developed has implications for clinical
genomics. By enabling accurate prediction of gene--disease
associations for novel diseases, our approach can improve the
prioritization of candidate genes in diagnostic settings. This can
potentially reduce the time and cost of rare disease diagnosis,
ultimately leading to earlier and more effective treatment
interventions.

\section*{Competing interests}
No competing interest is declared.

\section*{Author contributions statement}

FZC designed and conducted experiments, wrote the software, analyzed
and interpreted the results, and wrote the paper. R.H. conceived of
and supervised the work, acquired funding, and contributed to writing
of the manuscript. All authors have read and critically revised the
manuscript.

\section*{Acknowledgements}
We thank the MSc student at KAUST whose thesis research
\cite{safana-thesis} provided an early proof of concept for the
feasibility of \method{}, and whose work we extend.

\section*{Funding}

This work has been supported by funding from King Abdullah University
of Science and Technology (KAUST) Office of Sponsored Research (OSR)
under Award No. URF/1/4675-01-01, URF/1/4697-01-01, URF/1/5041-01-01,
REI/1/5235-01-01, and REI/1/5334-01-01.  This work was supported by
funding from King Abdullah University of Science and Technology
(KAUST) -- KAUST Center of Excellence for Smart Health (KCSH), under
award number 5932, and by funding from King Abdullah University of
Science and Technology (KAUST) -- Center of Excellence for Generative
AI, under award number 5940.


\bibliographystyle{natbib}
\bibliography{reference}






\end{document}